\documentclass[preprint,12pt]{elsarticle}
\usepackage{amsmath}
\usepackage{amsfonts}
\usepackage{bm}
\usepackage{lipsum}
\usepackage{enumerate}
\usepackage{graphicx}
\usepackage{amsfonts}

\usepackage{mathrsfs}
\usepackage{subfigure}
\usepackage{epstopdf}
\usepackage{url}
\usepackage{verbatim}
\usepackage{amssymb}
\usepackage{hyperref}
\usepackage{color}
\usepackage{listings}

\def\@begintheorem#1#2{\par\bgroup{\scshape #1\ #2. }\it\ignorespaces}
\def\@opargbegintheorem#1#2#3{\par\bgroup%
   {\scshape #1\ #2\ ({\upshape #3}). }\it\ignorespaces}
\def\@endtheorem{\egroup}

\if@onethmnum
  \newtheorem{theorem}{Theorem}
  \newtheorem{lemma}[theorem]{Lemma}
  \newtheorem{corollary}[theorem]{Corollary}
  \newtheorem{proposition}[theorem]{Proposition}
  \newtheorem{definition}[theorem]{Definition}
\newtheorem{example}[theorem]{Example}
\newtheorem{remark}[theorem]{Remark}
\newtheorem{homework}[theorem]{Homework}
\newtheorem{case}[theorem]{}

\else

\fi

\usepackage[T1]{fontenc}

\journal{Physica D (Nonlinear Phenomena) }

\begin{document}
%\maketitle
%\tableofcontents

\begin{frontmatter}



  % \title{Determinism in random flows:  an ergodic system for understanding noise propagation}
  \title{Determinism and invariant measures for diffusing passive scalars advected by unsteady random shear flows}
  

\author[1]{Lingyun Ding}
\ead{dingly@live.unc.edu}
\author[1]{Richard M. McLaughlin \corref{mycorrespondingauthor}}
\cortext[mycorrespondingauthor]{Corresponding author}
\ead{rmm@email.unc.edu}
\address[1]{Department of Mathematics, University of North Carolina, Chapel Hill, NC, 27599, United States}

%\ead{zctang@email.unc.edu}

 %\cortext[cor2]{Corresponding author}
 %\fntext[fn1]{This is the first author footnote.}
 %\fntext[fn2]{Another author footnote, this is a very long
 %  footnote and it should be a really long footnote. But this
 %  footnote is not yet sufficiently long enough to make two
 %  lines of footnote text.}
 %\fntext[fn3]{Yet another author footnote.}

% \address[1]{Department of Mathematics, the University of North Carolina at Chapel Hill}
% \address[2]{Elsevier B.V., Radarweg 29, 1043 NX Amsterdam,    The Netherlands}

%\ead[Label1]{}
% \address{Address\fnref{label1}}

\begin{abstract}
Here we study the long time behavior of an advection-diffusion equation
with a general time varying (including random) shear flow imposing no-flux boundary conditions on channel walls.  We derive the asymptotic approximation of the scalar field at long times by  using center manifold theory. We carefully compare it with existing time varying homogenization theory as well as other existing center manifold based studies, and present conditions on the flows under which our new approximations give a substantial improvement to these existing theories.  A recent study \cite{ding2020ergodicity} has shown  that Gaussian random shear flows induce a deterministic effective diffusivity at long times, and explicitly calculated the invariant measure. Here, with our established asymptotic expansions, we not only concisely demonstrate those prior conclusions for Gaussian random shear flows, but also generalize the conclusions regarding determinism to a much broader class of random (non-Gaussian) shear flows.   
 Such results are important ergodicity-like results in that they
assure an experimentalist need only perform a single realization of a
random flow to observe the ensemble moment predictions at long
time. Monte-Carlo simulations are presented illustrating how the
highly random behavior converges to the deterministic limit at long
time. Counterintuitively, we present a case demonstrating that the random flow may not induce larger dispersion than its deterministic counterpart, and in turn present rigorous conditions under which a random renewing flow induces a stronger effective diffusivity.
\end{abstract}



% 82C70Transport processes
% 82C80 Numerical methods (Monte Carlo, series resummation, etc.)
%34E13 Multiple scale methods
%76R50 diffusion
%37A25: Ergodicity, mixing, rates of mixing
%37H10: Generation, random and stochastic difference and differential equations
%37N10: Dynamical systems in fluid mechanics, oceanography and meteorology
\end{frontmatter}

\section{Introduction}
An extremely important class of problems concerns how fluid motion transports a diffusing scalar.
Since G. I. Taylor \cite{taylor1953dispersion} first introduced the calculation showing that a steady pressure driven flow in a pipe leads to a greatly enhanced effective diffusivity, the literature on this topic has exploded in many directions spanning many disciplines. Shortly following G. I. Taylor, Aris \cite{aris1956dispersion} presented an alternative approach for shear layers yielding a hierarchy for the spatial moments of the scalar field.  More recent moment analysis shows how the boundary geometry of the pipe can be used to control the distribution of solute which is  advected by the pressure driven flow \cite{aminian2016boundaries,aminian2015squaring,aminian2018mass}.

Unsteady flows typically generate different properties than their steady counterparts. Practical examples of unsteady flow include pulsatile blood flows \cite{salerno2020aris} and tidal estuaries \cite{chatwin1985mathematical}.  The first investigation of the Taylor dispersion in time-dependent flow dates back to Aris \cite{aris1960dispersion}, who presented the study of a solute advected by pulsating flow in a circular tube. After that, based on the Aris' moment method, a number of studies reported on the enhanced diffusivity induced by the single-frequency pulsating flow\cite{bowden1965horizontal,chatwin1975longitudinal,watson1983diffusion,mukherjee1988dispersion,jimenez1984contaminant}, the single frequency Couette-Poiseuille \cite{bandyopadhyay1999contaminant,paul2008dispersion,barik2019multi,barik2017transport} and the multi-frequency flow \cite{vedel2012transient,vedel2014time,ding2021enhanced}.  Alternative approaches, using center manifold theory, \cite{mercer1990centre,mercer1994complete,marbach2019active,masri2019reduced} or homogenization methods \cite{chu2019dispersion,chu2020dispersion} not only predict the effective diffusivity but also gives the direct expression for the full concentration field at long times.

We notice that three points haven't been addressed well in the literature regarding to the shear dispersion in time-varying flows. First, most of those theoretical studies focused on the cross-sectional averaged concentration, while fewer studies have explored asymptotic corrections which capture cross-channel variations. Here, with the center manifold theory, we present a systematic procedure to construct an approximation to capture the traverse variation of the scalar field. Second, several interesting articles \cite{mercer1990centre,mercer1994complete,marbach2019active} implemented center manifold theory for such unsteady problems employing certain slowly varying assumptions to simplify the calculation. Such assumptions restrict the applicability of the effective dynamics. Here we relax this assumption by carefully incorporating the temporal fluctuation of the flows into the analysis. Hence, our results can handle rapidly fluctuating flows or even random flows.
Third, recent results have explicitly calculated using statistical moment closure the invariant measure for a diffusing passive scalar advected by a class of random shear flows \cite{ding2020ergodicity,camassa2021persisting} employing no-flux boundary conditions on channel domains.  These results generalize prior turbulent intermittency in free space of Majda \cite{kraichnan1968small} and Kraichnan \cite{majda1993random}. 
Interestingly, we establish here how center manifold theory can be used to greatly extend these theories to a much broader class of random shear flows, particularly regarding their temporal statistics.  In doing so, we can extend results which show how all the effective diffusion coefficients converge to a deterministic value for this broader class of flows, in sharp contrast to the free-space analog considered by Majda and others \cite{mclaughlin1996explicit,bronski1997scalar} in which the effective diffusivity is random at all times.  Such results are important ergodicity-like results in that they assure an experimentalist need only perform a single realization of a random flow to observe the field moment predictions at long time.  

The paper is organized as follows. In section~\ref{sec:setup}, we formulate the governing equation of the shear dispersion problem and review the Aris moment method. In section \ref{sec:centerManifold}, we discuss the procedure of applying center manifold theory to the Taylor dispersion problem with time-varying shear flow. By utilizing the first-order approximation of the cross-sectional averaged concentration, we present a nonnegative asymptotic expansion of the scalar field at long times which captures the transverse variations. We document situations in which a time varying cell problem produces a more accurate approximation than the parametric (adiabatic) approach employed recently \cite{mercer1990centre,mercer1994complete,marbach2019active}.  In section \ref{sec:randomflow}, we demonstrate that a class of flows with finite correlation time will induce a deterministic effective diffusivity at long times. Moreover, we establish conditions which guarantee that the periodic in time problem always yields a weaker effective diffusivity than the random counterpart.  With the derived effective equation, we computed the explicit formula of invariant measure of the random passive field.

\section{Setup and background of the problem}\label{sec:setup}
\subsection{Governing Equation and Nondimensionalization}

\subsubsection{ Advection-diffusion Equation}
We consider the problem in a channel domain $(x, \mathbf{y}) \in \mathbb{R}\times \Omega$, where the $x$-direction is the longitudinal direction of the channel and $\Omega \subset \mathbb{R}^{d}$ stands for the cross section of the channel. Some practical examples of the boundary geometry includes the parallel-plate channel $\Omega= \left\{ y| y \in [0,L]  \right\}$, the circular pipe $\Omega=\left\{ \mathbf{y}| \mathbf{y}^{2}\leq L \right\}$, the rectangular duct $\Omega=\left\{ \mathbf{y}| \mathbf{y} \in [0,L]^{2} \right\}$, bowed rectangular channels \cite{lee2021dispersion}. 
The passive scalar is governed by the advection-diffusion equation with  a general time-varying shear flow $u(\mathbf{y},t)$ and no-flux boundary condition which takes the form
\begin{equation}\label{eq:AdvectionDiffusionEquation}
\partial_{t}T+ v(\mathbf{y},t) \partial_{x}T=  \kappa \Delta T, \quad T(x,\mathbf{y},0)= T_{I}(x,\mathbf{y}),\quad  \left. \frac{\partial T}{\partial \mathbf{n}} \right|_{\mathbb{R}\times \partial \Omega}=  0,
\end{equation}
where $\kappa$ is the diffusivity, $T_{I}(x,\mathbf{y})$ is the initial data, $\mathbf{n}$ is the outward normal vector of the boundary $\mathbb{R} \times \partial\Omega$ and $\partial\Omega$ is the boundary of $\Omega$.

\subsubsection{Nondimensionalization}
With the change of variables
\begin{equation}
\begin{aligned}
&Lx'=x, \quad  L\mathbf{y}'=\mathbf{y}, \quad \frac{L^2}{\kappa} t'=t, \quad  Uv' (\mathbf{y}',t')= v (\mathbf{y},t),\quad L\Omega'=\Omega,\\
&T_I' (x',\mathbf{y}') L^{-d-1}\int\limits_{\mathbb{R}\times \Omega}^{}T_{I}(x,\mathbf{y})\mathrm{d}x\mathrm{d}\mathbf{y} =T_{I} (x,\mathbf{y}), \\
&T' (x',\mathbf{y}',t') L^{-d-1}\int\limits_{\mathbb{R}\times \Omega}^{}T_{I} (x,\mathbf{y})\mathrm{d}x\mathrm{d}\mathbf{y} =T (x,\mathbf{y},t), \;
\end{aligned}
\end{equation}
after dropping the primes, we obtain the nondimensionalized advection-diffusion equation
\begin{equation}\label{eq:AdvectionDiffusionEquationNon}
\partial_{t}T+ \mathrm{Pe}v(\mathbf{y},t) \partial_{x}T=  \Delta T,\; T(x,\mathbf{y},0)= T_{I}(x,\mathbf{y}),\; \left. \frac{\partial T}{\partial \mathbf{n}} \right|_{\mathbb{R} \times \partial \Omega}=  0, 
\end{equation}
where $\mathrm{Pe}=  {U L}/{ \kappa}$ is the P\'{e}clet number.

\subsection{Aris Moment Hierarchy}
Aris showed in \cite{aris1956dispersion} that one could write down a recursive system of partial differential equations \eqref{eq:AdvectionDiffusionEquationNon} for the spatial moments of the tracer $T$. 
The $n$th Aris moment is defined by $T_n (\mathbf{y},t) = \int\limits_{-\infty}^{\infty} x^n T(x,\mathbf{y},t) \mathrm{d} x$. With the assumption $T(\pm \infty, \mathbf{y}, t)=0$, the Aris moments satisfy the recursive relationship called the Aris equations,
\begin{equation}\label{eq:ArisMomentDef}
\begin{aligned}
&(\partial_t- \Delta)T_n=   n (n-1)T_{n-2}+ n  \mathrm{Pe}v(\mathbf{y},t) T_{n-1},\;
T_n(\mathbf{y},0)= \int\limits_{-\infty}^{\infty} x^n T_{I}(x,\mathbf{y}) \mathrm{d} x, \;
\left. \frac{\partial T}{\partial \mathbf{n}} \right|_{\partial \Omega}=  0, \\
\end{aligned}
\end{equation}
where $T_{n}=0$ if $n\leq -1$. The full moments of $T$ are then obtained though the cross-sectional average of the moments $ \bar{T}_n = \frac{1}{\left| \Omega\right|}  \int\limits_{\Omega}^{}T_n\mathrm{d}\mathbf{y}$, where  $\left| \Omega \right|$ is the area of $\Omega$. In this following context, we use the overline to denote the cross sectional average. Applying the divergence theorem and boundary conditions yield
\begin{equation}\label{eq:averArisMomentDef}
\begin{aligned}
&\frac{\mathrm{d}  \bar{T}_{n} }{\mathrm{d} t}=  n(n-1)  \bar{T}_{n-2}+ n  \mathrm{Pe}  \overline{v(\mathbf{y},t)  T_{n-1}} ,\;  \bar{T}_n (0)=\frac{1}{\left| \Omega \right|} \int\limits_{\Omega}^{}\int\limits_{-\infty}^{\infty} x^n T_{I}(x,\mathbf{y}) \mathrm{d} x \mathrm{d}\mathbf{y}.\\
\end{aligned}
\end{equation}
The homogenization method in \cite{ding2021enhanced,camassa2010exact}  suggests that, assuming a scale separation in the initial data, the solution of equation \eqref{eq:AdvectionDiffusionEquation} can be approximated by a diffusion equation with an effective diffusion coefficient. The effective longitudinal effective diffusivity could be computed through the Aris moments
\begin{equation}\label{eq:effectiveDiffusivityDefinition}
\kappa_{\mathrm{eff}}=   \lim\limits_{t\rightarrow \infty}\frac{\mathrm{Var} ( \bar{T})}{2t},
\end{equation}
where $\mathrm{Var} (\bar{T})= \bar{T}_2 - \bar{T}_1^2$ is the variance of the cross-sectional average $ \bar{T}$. In this paper, we use $\kappa_{\mathrm{eff}}$ to denote the dimensional effective diffusivity computed by the dimensional Aris moment and  use the $\tilde{\kappa}_{\mathrm{eff}}=\kappa_{\mathrm{eff}}/\kappa$ to denote the non-dimensional effective diffusivity.

The effective diffusivity characterizes the symmetric property of the longitudinal distribution. We are also interested in the asymmetry properties of $\bar{T}$. Skewness is the lowest order integral measure of the asymmetry of a real-valued probability distribution, which is defined as
\begin{equation}\label{eq:SkewnessDefinition}
  \mathrm{S} (\bar{T})=   \frac{ \bar{T}_3- 3 \bar{T}_{2} \bar{T}_{1}+2 \bar{T}_{1}^{3}}
  {\left(\bar{T}_2 - \bar{T}_1^2\right)^{\frac{3}{2}}}.
\end{equation}
The information of shape provided by the skewness could improve the design of microfluidic flow injection analysis \cite{aminian2018mass,trojanowicz2016recent} and chromatographic separation \cite{blom2003chip}.

\section{ Center manifold description of the shear dispersion problem}\label{sec:centerManifold}
\subsection{Center manifold and reduction principle}
 In pioneering work, Mercer and Roberts \cite{mercer1990centre} interpreted the long time asymptotic of the shear dispersion problem as  the center manifold of a dynamical system, which provides a systematic and near rigorous approach to derive the approximation. Besides the shear dispersion problem, the pratical applications of center manifold theory include chromatographic model and reactors\cite{balakotaiah1995dispersion}, elastic beam deformations\cite{roberts1993invariant}, and  thin fluid flows dynamics \cite{roberts1996low,roberts2006accurate}.
To explain the center manifold method, let's consider an autonomous differential system of the form
\begin{equation}\label{eq:dynamicalSystem1}
\begin{aligned}
&\frac{\mathrm{d} \mathbf{x}}{\mathrm{d} t}= A \mathbf{x}+ f(\mathbf{x},\mathbf{y}),\quad \frac{\mathrm{d} \mathbf{y}}{\mathrm{d} t}= B \mathbf{y}+ g(\mathbf{x},\mathbf{y}), \\
\end{aligned}
\end{equation}
where $\mathbf{x}\in \mathbb{R}^m,\mathbf{y}\in \mathbb{R}^{n}$. $A,B$ are matrices whose eigenvalues have vanishing and negative real parts, respectively. $f(\mathbf{x},\mathbf{y})$, $g (\mathbf{x},\mathbf{y})$ and their first order partial derivatives are zero at $\mathbf{x}=\mathbf{0},\mathbf{y}=\mathbf{0}$. These conditions grantee the existence of a center manifold $\mathbf{y}=h(\mathbf{x})$ which has two important features. First, the stability properties of the dynamical system \eqref{eq:dynamicalSystem1} at the origin are shared by the following lower dimensional equation
\begin{equation}\label{eq:dynamicalSystem2}
\begin{aligned}
\frac{\mathrm{d} \mathbf{x}}{\mathrm{d} t}= A \mathbf{x}+ f(\mathbf{x},h(\mathbf{x})).
\end{aligned}
\end{equation}
Second, in case of a stable equilibrium $(\mathbf{x},\mathbf{y})= (\mathbf{0},\mathbf{0})$ each solution of system \eqref{eq:dynamicalSystem1} which starts close to the origin exponentially decays to a particular solution on the center manifold \cite{carr2012applications,carr1983application}. With these two features of the center manifold, one can reduce the original $m+n$-dimensional system \eqref{eq:dynamicalSystem1} to a $m$-dimensional system \eqref{eq:dynamicalSystem2} with only the price of  exponential corrections.

This classical center manifold theory and reduction principle could be generalized in many directions. First, the dynamical system \eqref{eq:dynamicalSystem1} could be an infinite dimensional system where the matrices $A$,$B$ become linear operators \cite{carr1983application2}. Second, similar results hold for a more general dynamical system $\frac{\mathrm{d} \mathbf{x}_{i}}{\mathrm{d} t}=A_{i}\mathbf{x}_{i}+ f (\mathbf{x}_{1},...,\mathbf{x}_{N},t)$, $1\leq i\leq N$ and the restriction of eigenvalues could be weakened \cite{van1993reduction,aulbach1996integral}.  This generalization leads to a so-called two-mode invariant manifold model for the shear dispersion problem \cite{watt1995accurate,watt1996construction,smith1987diffusion}. Third, more related to our topic, the system could be non-autonomous, where the center manifold becomes time-dependent $\mathbf{y}=h (\mathbf{x},t)$ \cite{aulbach1982reduction,aulbach1999invariant,roberts2018backwards}. For further details regarding center manifold theory, we refer to \cite{carr2012applications,aulbach1996integral}  and references therein.

Notice that the advection-diffusion equation \eqref{eq:AdvectionDiffusionEquationNon} is linear, while the center manifold theory applies to a system with nonlinear terms. To fit the center manifold theory, we first apply the Fourier transform on equation \eqref{eq:AdvectionDiffusionEquationNon} and obtain
\begin{equation}\label{eq:AdvectionDiffusionEquationNonFourier}
\begin{aligned}
 &\frac{\partial \hat{T}}{\partial t}-\mathrm{i} k \mathrm{Pe}v(\mathbf{y},t)  \hat{T}= -k^{2}\hat{T}+ \Delta_{\mathbf{y}}\hat{T},\;
 \left. \frac{\partial \hat{T}}{\partial \mathbf{n}} \right|_{\mathbb{R} \times \partial \Omega}=  0,\; \hat{T}(k,\mathbf{y},0)= \hat{T}_{I}(k,\mathbf{y}), \\
\end{aligned}
\end{equation}
Second, we conceptually non-linearize equation \eqref{eq:AdvectionDiffusionEquationNonFourier} by treating the wavenumber as a dependent variable of the dynamical system. Notice that $\Delta_{\mathbf{y}}$ has a null space which consists of all function independent on $y$. To fit the form of equation \eqref{eq:dynamicalSystem1}, we rewrite equation \eqref{eq:AdvectionDiffusionEquationNonFourier} as
\begin{equation}
\begin{aligned}
&\partial_{t}
\begin{bmatrix}
  k\\
 \hat{\overline{T}}
\end{bmatrix}
 =
 \begin{bmatrix}
   0&0\\
   0& \Delta_{\mathbf{y}}
 \end{bmatrix}
\begin{bmatrix}
  k\\
 \hat{\overline{T}}
\end{bmatrix}
+
\begin{bmatrix}
  0\\
  \mathrm{i} k \mathrm{Pe}\overline{v(\mathbf{y},t)  \hat{T}}-k^{2}\hat{\overline{T}}
\end{bmatrix}
,\\
&\hat{T'}= \Delta_{\mathbf{y}}\hat{T'}+\mathrm{i} k \mathrm{Pe}v(\mathbf{y},t)  \hat{T'}-k^{2}\hat{T}-\overline{\mathrm{i} k \mathrm{Pe}v(\mathbf{y},t)  \hat{T}},\\
& \left. \frac{\partial \hat{T}}{\partial \mathbf{n}} \right|_{\mathbb{R} \times \partial \Omega}=  0,\quad \hat{T}(k,\mathbf{y},0)= \hat{T}_{I}(k,\mathbf{y}),
\end{aligned}
\end{equation}
where $T' (x,\mathbf{y},t)$ and $\overline{T} (x,t)$ are the fluctuation and average of $T (x,\mathbf{y},t)$ with respect to  $\mathbf{y}$. This system admits a center manifold $\hat{T}'=h (\hat{\overline{T}},k,t)$. Based on center manifold theory, $\hat{T}$ converges to $h (\hat{\overline{T}},k,t)+\hat{\overline{T}}$ exponentially as $t\rightarrow \infty$. Due to the diffusion effect, $T$ is a decaying scalar field. The energy concentrates near the neighborhood of $k=0$ at long times. Hence, we can seek the expansion of $h (\hat{\overline{T}},k,t)$ for small $k$ and $\hat{\overline{T}}$, $h=\sum\limits_{n=1}^{\infty}h_{n} (\mathbf{y},t) k^{n}\hat{\overline{T}}+\mathcal{O} (\hat{\overline{T}}^{2})$. That is equivalent to approximating the scalar field $T$ by the derivatives of its cross-sectional average $\bar{T}$ with respect to $x$. This idea dated back to Gill \cite{gill1967note,gill1970exact} and also has been discussed in \cite{young1991shear}.

For simplicity, we rewrite all equations in term of physical variables. The governing equations are 
\begin{equation}\label{eq:ADEMeanFluctuation}
\begin{aligned}
\partial_{t}\bar{T}&=  \partial_{x}^{2}\bar{T}- \mathrm{Pe} \overline{v(\mathbf{y},t) \partial_{x} T},\\
\partial_{t}T&=  \Delta_{\mathbf{y}}T+\partial_{x}^{2}T- \mathrm{Pe}v(\mathbf{y},t) \partial_{x} T.\\
\end{aligned}
\end{equation}
The expansion becomes 
\begin{equation}\label{eq:ADECenterManifoldExpansion}
\begin{aligned}
T&=T'+\bar{T}=\bar{T}+h(\bar{T})=\sum\limits_{n=0}^{\infty} \theta_n (\mathbf{y},t) \partial_x^n \bar{T}. \\
\end{aligned}
\end{equation}
The fluctuation is mean zero, $\int_{\Omega}^{}T'\mathrm{d} \mathbf{y}=0$, which implies $\bar{\theta}_{0}=1$ and $\bar{\theta}_n=0$ if $n\geq 1$ at long times. We have $\left. \frac{\partial}{\partial \mathbf{n}} \theta_n \right|_{\mathbf{y}\in \partial\Omega}=0$ from the no-flux boundary conditions of $T$. Substituting expansion \eqref{eq:ADECenterManifoldExpansion} into equation \eqref{eq:ADEMeanFluctuation}, we have
\begin{subequations}
\begin{equation}\label{eq:ADEMeanEvolution}
  \begin{aligned}
& \partial_{t}\bar{T}=  \partial_{x}^{2}\bar{T}- \mathrm{Pe} \overline{v \partial_{x} T},\\
\end{aligned}
\end{equation}
\begin{equation}
  \begin{aligned}
& \sum\limits_{n=0}^{\infty} \partial_{t} \theta_n \partial_x^n \bar{T}+  \sum\limits_{n=0}^{\infty} \theta_n  \partial_x^n \partial_{t}\bar{T}= \Delta_{\mathbf{y}}T+\partial_{x}^{2}T- \mathrm{Pe}v \partial_{x} T.\\
\end{aligned}
\end{equation}  
\end{subequations}

Grouping all terms of the same order, namely $\partial_x^{n}\bar{T}$, we find that we have to solve the sequence of equations

\begin{equation}\label{eq:ADECenterManifoldCell}
\begin{aligned}
&\left( \partial_{t}-\Delta_{\mathbf{y}} \right) \theta_0 =0,\\
&\left( \partial_{t}-\Delta_{\mathbf{y}} \right) \theta_1= -\mathrm{Pe}\theta_0 \left( v-\overline{\theta_{0} v} \right),\\
&\left( \partial_{t}-\Delta_{\mathbf{y}} \right) \theta_n=-\mathrm{Pe}v \theta_{n-1}+\mathrm{Pe}\sum\limits_{m=0}^{n-1}\theta_{n-m-1} \overline{v \theta_{m}} ,
\end{aligned}
\end{equation}
where $\theta_n=0$ if $n<0$. 
After we solve $\theta_{n}$ successively, we obtain the closed evolution equation of $\bar{T}$ by substituting $T=\sum\limits_{n=0}^{\infty} \theta_n (\mathbf{y},t) \partial_x^n \bar{T}$ into equation \eqref{eq:ADEMeanFluctuation},
\begin{equation}\label{eq:ADECenterManifoldMean}
\begin{aligned}
\partial_{t}\bar{T}&=  \partial_{x}^{2}\bar{T}- \mathrm{Pe}  \sum\limits_{n=0}^{\infty}\overline{v\theta_{n} } \partial_{x}^{n+1} T.\\
\end{aligned}
\end{equation}
Finlay, once we solve equation \eqref{eq:ADECenterManifoldMean} for $\bar{T}$, we obtain the approximation of the scalar field $T$ via expansion \eqref{eq:ADECenterManifoldExpansion}. 

\subsection{The first and second order effective equation}
In this subsection, we will compute equation  \eqref{eq:ADECenterManifoldCell} and \eqref{eq:ADECenterManifoldMean}  for the flow $v (\mathbf{y}, t)= \xi (t)u (\mathbf{y})$. For more general non-separable flow $v (\mathbf{y},t)$, one could reduce it to a separable form by utilizing the Fourier transform in time.
To simplify the calculation, we assume $T_{I} (x,\mathbf{y})=\delta (x)$. Otherwise, the general initial condition only creates extra exponential decaying terms and yields the same asymptotic expansion at long times.

With the constraints of the average and boundary conditions of $\theta_{n}$, we have $\theta_0=1$. Therefore, the equation of $\theta_1$ becomes
\begin{equation}\label{eq:centerCellProblemOrder1}
\begin{aligned}
\left( \partial_{t}-\Delta_{\mathbf{y}} \right) \theta_1=-\mathrm{Pe} (v-\bar{v}),\quad \left. \frac{\partial}{\partial_{\mathbf{n}}}\theta_{1} \right|_{\partial \Omega }=0,\\
\end{aligned} 
\end{equation}
which is identical to  equation \eqref{eq:ArisMomentDef} in the Aris moments calculation. Since the theory concerns the long time dynamics of the scalar field and the long time limit of $\theta_1$ doesn't depend on the initial condition, in principle, one can solve equation \eqref{eq:centerCellProblemOrder1} with arbitrary initial condition. To obtain a better approximation at earlier stage, one can choose suitable initial condition of $\theta_n$ to match both sides of the expansion \eqref{eq:ADECenterManifoldExpansion} at $t=0$.  Then when $v (\mathbf{y}, t)= \xi (t)u (\mathbf{y})$, the solution of equation \eqref{eq:centerCellProblemOrder1} is
\begin{equation}\label{eq:centerCellProblemOrder1expansion1}
  \begin{aligned}
    \theta_1 (\mathbf{y},t) =  -\mathrm{Pe}\sum\limits_{n=1}^{\infty}   \phi_{n} \left\langle u, \phi_{n}\right\rangle  \int _0^te^{\lambda_{n} (s-t)} \xi(s)ds. \\
\end{aligned}
\end{equation}
where $\left\langle f,g \right\rangle=\frac{1}{|\Omega|} \int\limits_{\Omega}^{}fg \mathrm{d} \mathbf{y}$.  $\phi_{n}, \lambda_{n}$ are the eigenfunctions and eigenvalues of the Laplace operator in the cross section of the channel $\Omega$ with no-flux boundary condition, i.e.,
\begin{equation}
\begin{aligned}
-\Delta \phi_{n}=\lambda_{n}\phi_{n}, \quad \left. \frac{\partial}{\partial_{\mathbf{n}}}\phi_{n} \right|_{\partial \Omega }=0, \quad \left\langle \phi_n, \phi_n \right\rangle=1. \\
\end{aligned}
\end{equation}

Substituting $T=\bar{T}+\theta_1\partial_x\bar{T}$ into the evolution equation of $\bar{T}$, we obtain the first order effective equation
\begin{equation}\label{eq:centerEffectiveOrder1}
\begin{aligned}
\partial_{t}\bar{T}+\mathrm{Pe}\bar{v}\partial_{x} \bar{T}=  a_{2}\partial_{x}^{2}\bar{T}, \quad  a_{2}=  \left( 1-\mathrm{Pe}\overline{v\theta_{1} } \right). \\
\end{aligned}
\end{equation}

The classical homogenization approach relies on the Fredholm alternative which involves a space-time average. As a result, the effective equation is a constant coefficient equation even for the time-varying flow case \cite{ding2021enhanced,chu2019dispersion}. Here, with the center manifold approach, we obtain the effective equation \eqref{eq:centerEffectiveOrder1} with time-dependent coefficients which could approximate the scalar field better in an earlier stage. Comparing with the definition of Aris moments and  variance of the cross-sectional average, we have
\begin{equation}
\begin{aligned}
\mathrm{Var} ( \bar{T})=\mathrm{Var} ( \bar{T_{I}})+ 2\int\limits_0^ta_{2} (s)\mathrm{d}s
\end{aligned}
\end{equation}
For a periodic time-varying flow \cite{vedel2012transient} and a class of random flows \cite{camassa2021persisting,ding2020ergodicity}, we have $\mathrm{Var} ( \bar{T})= 2 \kappa_{\mathrm{eff}}t+ \mathcal{O} (1)$, where $\kappa_{\mathrm{eff}}$ is the effective diffusivity
\begin{equation}
\begin{aligned}
& \quad \kappa_{\mathrm{eff}}= \lim\limits_{t\rightarrow \infty}\frac{\mathrm{Var} ( \bar{T})}{2t}= \lim\limits_{t\rightarrow \infty} \frac{1}{t} \int\limits_0^t a_{2} (s)\mathrm{d} s. \\
\end{aligned}
\end{equation}
In other words, $a_2$ can be approximated by its time average at long times.

With the expression  \eqref{eq:centerCellProblemOrder1expansion1}, the effective diffusivity induced by the flow $v (\mathbf{y}, t)= \xi (t)u (\mathbf{y})$ is
\begin{equation}\label{eq:centeManifoldrEffectiveDiffusivity}
  \begin{aligned}
 &a_{2}=  \left( 1-\mathrm{Pe}\overline{v\theta_{1} } \right)= 1+\mathrm{Pe}^{2} \sum\limits_{n=1}^{\infty} \left\langle u,\phi_{n} \right\rangle^{2}\xi (t)\int\limits _0^te^{\lambda_{n} (s-t)} \xi(s) \mathrm{d} s,\\
&\kappa_{\mathrm{eff}}=1+ \mathrm{Pe}^{2} \lim\limits_{t\rightarrow \infty} \frac{1}{t} \int\limits_0^t \sum\limits_{n=1}^{\infty} \left\langle u,\phi_{n} \right\rangle^{2}\xi (s_{2})\int\limits _0^{s_{2}}e^{\lambda_{n} (s_{1}-s_{2})} \xi(s_{1}) \mathrm{d} s_{1} \mathrm{d} s_{2}.
\end{aligned}
\end{equation}

\begin{figure}
  \centering
    \includegraphics[width=0.46\linewidth]{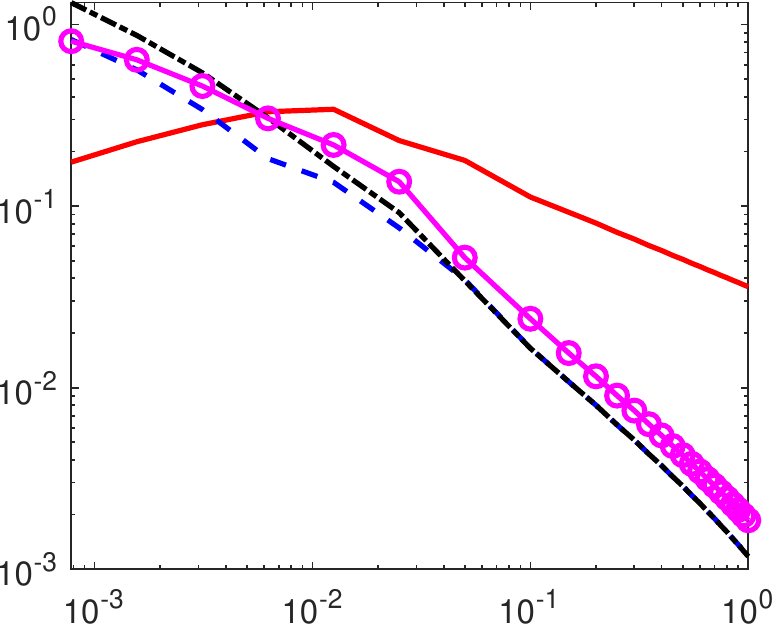}
  \hfill
  \caption[]
  {The relative errors of different approximations for the solution studied in right panel of figure \ref{fig:plotRoberts}. The red solid curve, blue dashed curve, black dashed dot curve and purple curve are the relative error of approximation \eqref{eq:approximationTo1v0}, \eqref{eq:approximationTo1v1}, \eqref{eq:approximationTo1v3} and \eqref{eq:approximationTo1v2}, respectively.  }
  \label{fig:nonnegativeApproximation}
\end{figure}

With the initial condition $T (x,\mathbf{y},0)=\delta (x)$, the solution of equation \eqref{eq:centerEffectiveOrder1} gives an approximation of $\bar{T}$ as $t\rightarrow \infty$,
\begin{equation}\label{eq:approximationTo1v0}
\begin{aligned}
&  \bar{T} (x,t) =  \frac{1}{\sqrt{4\pi b_{2} }} \exp \left( \frac{-\tilde{x}^{2}}{ 4 b_{2}} \right) + \mathcal{O} (t^{-\frac{3}{2}}),\\
  &b_{2}=\int\limits_0^t a_{2} (s)\mathrm{d}s,\quad  \tilde{x}=x- \mathrm{Pe}\int\limits_0^t \bar{v} (s)\mathrm{d} s. \\
\end{aligned}
\end{equation}
For steady flow, we have $b_{2}= \kappa_{\mathrm{eff}}t$. Then equation \eqref{eq:approximationTo1v0} reduces to the classical Gaussian approximation \cite{chatwin1970approach}.
Since the scalar field will be homogenized across the channel at long times, $\bar{T}$ itself could be an approximation of $T$. In fact, we could obtain a more accurate approximation of $T$, 
\begin{equation}\label{eq:approximationTo1v1}
\begin{aligned}
  T\approx \bar{T}+\theta_1\partial_x\bar{T}= \left( 1- \frac{\theta_{1} (\mathbf{y},t)\tilde{x}}{2 b_{2}}\right)\frac{1}{\sqrt{4\pi b_{2}}} \exp \left( \frac{-\tilde{x}^{2}}{4 b_{2}} \right)+ \mathcal{O} (t^{-\frac{3}{2}}).
\end{aligned}
\end{equation}
Since $\partial_{x}\bar{T}$ is an odd function with respect to $x$, the error of approximation \eqref{eq:approximationTo1v1} is still $\mathcal{O} (t^{-\frac{3}{2}})$. However, equation \eqref{eq:approximationTo1v1} practically performs better than $\bar{T}$ because it depicts the across channel variation of the scalar. Figure \ref{fig:nonnegativeApproximation} presents the relative errors of different approximations for the solution studied in right panel of figure \ref{fig:plotRoberts}, where the metric of error is ${\lVert T-T_{approx}\rVert _{\infty}}/{\lVert T\rVert _{\infty}}$. 
As shown in figure \ref{fig:nonnegativeApproximation}, the relative error of approximation \eqref{eq:approximationTo1v0} (red curve) is around $0.1$ at $t=1$, while, the relative error of approximation \eqref{eq:approximationTo1v1} (blue curve) is around $10^{-3}$.  Since two approximations are of the same asymptotic order at long times, presumably the differences between the two approximations will reduce as time is further increased.  

In many applications, the scalar field usually stands for the concentration which must be nonnegative for all times.  However, this approximation \eqref{eq:approximationTo1v1} could be negative for some $x$ and $t$, which may not be desirable in those applications. \cite{wu2016spatial} proposed the following nonnegative approximation to study the  transverse distribution of  concentration distribution for laminar tube flow,
\begin{equation}\label{eq:approximationTo1v2}
\begin{aligned}
T\approx & \frac{1}{\sqrt{4\pi b_{2}}} \exp \left( \frac{-(\tilde{x}-\theta_{1} (\mathbf{y},t))^{2}}{4 b_{2}} \right). \\
\end{aligned}
\end{equation}
The approximation \eqref{eq:approximationTo1v2} converges asymptotically to approximation \eqref{eq:approximationTo1v1} as $\theta\rightarrow 0$. However, the relative difference between them doesn't vanish as $t\rightarrow \infty$. As shown in figure \ref{fig:nonnegativeApproximation}, there is a visible difference between the approximation \eqref{eq:approximationTo1v2} (purple curve) and \eqref{eq:approximationTo1v1} (blue curve).

Here, we propose a nonnegative asymptotic expansion 
\begin{equation}\label{eq:approximationTo1v3}
\begin{aligned}
T\approx & \left( 1- \frac{\theta_{1} (\mathbf{y},t)\tilde{x}}{4 b_{2}} \right)^{2}\frac{1}{\sqrt{4\pi b_{2}}} \exp \left( \frac{-\tilde{x}^{2}}{4 b_{2}} \right), \quad t\rightarrow \infty. \\
\end{aligned} 
\end{equation}
Since the difference between equation \eqref{eq:approximationTo1v1} and \eqref{eq:approximationTo1v3} is $\mathcal{O} (t^{-\frac{5}{2}})$, the relative difference between them vanishes as $t\rightarrow \infty$. From figure \ref{fig:nonnegativeApproximation}, we can see that the relative error of approximation \eqref{eq:approximationTo1v1} and \eqref{eq:approximationTo1v3} is almost indistinguishable after $t=0.1$.

Next, we study the second order approximation of the scalar field. We have to solve the equation for $\theta_{2}$,
\begin{equation}\label{eq:centerCellProblemOrder2}
\begin{aligned}
&\left( \partial_{t}-\Delta_{\mathbf{y}} \right) \theta_2= -\mathrm{Pe}(v \theta_{1}-\theta_{1}\bar{v}- \overline{v \theta_{1}}),\quad \left. \frac{\partial}{\partial_{\mathbf{n}}}\theta_{2} \right|_{\partial \Omega }=0, 
\end{aligned}
\end{equation}
We have the expansion of $v \theta_{1}-\theta_{1}\bar{v}- \overline{v \theta_{1}}$,
\begin{equation}
\begin{aligned}
v \theta_{1}-\theta_{1}\bar{v}- \overline{v \theta_{1}}= & \sum\limits_{n_2,n_{1}=1}^{\infty}\left\langle \theta_{1},\phi_{n_{1}} \right\rangle \left\langle \phi_{n_{1}}(v-\bar{v}), \phi_{n_2} \right\rangle \phi_{n_2}.\\
\end{aligned}
\end{equation}
That leads to the solution
\begin{equation}
\begin{aligned}
  \theta_2=  &\mathrm{Pe}^{2}\sum\limits_{n_2,n_{1}=1}^{\infty} \left\langle u, \phi_{n_{1}}\right\rangle \left\langle \phi_{n_{1}}(u-\bar{u}), \phi_{n_2} \right\rangle \phi_{n_2} \int\limits_0^t \left( e^{\lambda_{n_{2}} (s_{2}-t)}  \xi(s_{2})  \int _0^{s_{2}}e^{\lambda_{n_{1}} (s_{1}-s_{2})} \xi(s_{1})ds_{1} \right) \mathrm{d} s_{2}  \\  
\end{aligned}
\end{equation}
Substituting $T=\bar{T}+\theta_1\partial_x\bar{T}+\theta_{2}\partial^{2}_{x}\bar{T}$ into the evolution equation of $\bar{T}$, the approximated evolution equation for $\bar{T}$ becomes a linearized Burgers-Korteweg-de Vries equation
\begin{equation}
\begin{aligned}
  &  \partial_{t}\bar{T}+ \bar{v} \partial_{x}\bar{T}= a_{2}\partial_{x}^{2}\bar{T} - a_{3} \partial_{x}^{3} \bar{T}, \quad a_{3}=\mathrm{Pe}\overline{ v \theta_{2}}.
\end{aligned}
\end{equation}
Then we can consider two cases based on $a_{3}=\mathrm{Pe}\overline{ v \theta_{2}}$. First, we consider the case in involving $a_{3}=\mathrm{Pe}\overline{ v \theta_{2}}=0$, which implies the skewness of $\bar{T}$ is zero. One such example is the linear shear flow created by moving one boundary of parallel-plate channel \cite{ding2021enhanced}. In this case, the evolution equation for $\bar{T}$ reduces to the diffusion equation, where the Gaussian approximation  \eqref{eq:approximationTo1v0} is still valid. Then we obtain the approximation of the whole scalar field
\begin{equation}\label{eq:approximationTo2v1}
\begin{aligned}
T&=\bar{T}+\theta_1\partial_x\bar{T}+\theta_{2}\partial^{2}_{x}\bar{T} \\
  &=\left( 1- \frac{\theta_{1} \tilde{x}}{2 b_{2}}+\frac{\theta _2 \left(\tilde{x}^2-2 b_2\right)}{4 b_2^2}\right)\frac{1}{\sqrt{4\pi b_{2}}} \exp \left( \frac{-\tilde{x}^{2}}{4 b_{2}} \right)+ \mathcal{O} (t^{-2}). \\
\end{aligned}
\end{equation}
Since $\partial_{x}^{2}\bar{T}$ is an even function with respect to $x$, the error of approximation \eqref{eq:approximationTo2v1} is $\mathcal{O} (t^{-2})$ which is more accurate than the approximation \eqref{eq:approximationTo1v1}.

\begin{figure}
  \centering
  \subfigure[]{
    \includegraphics[width=0.46\linewidth]{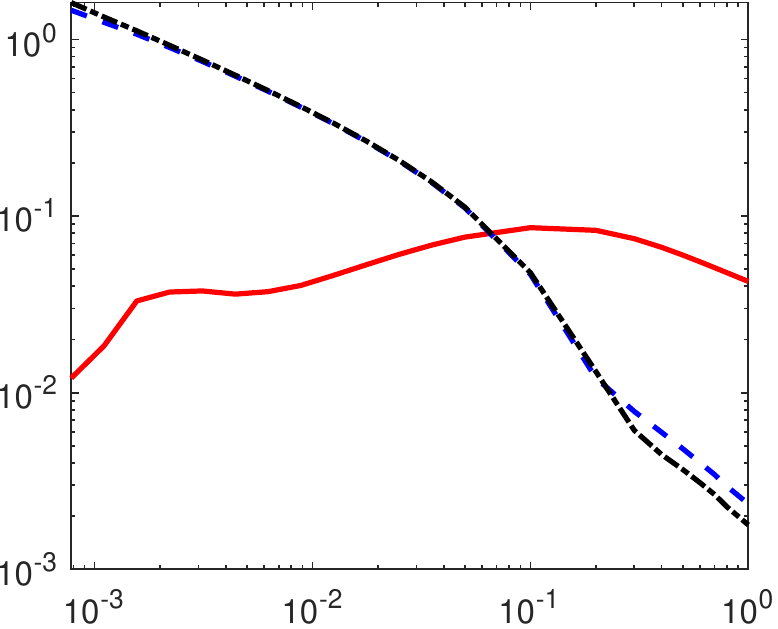}
  }
    \subfigure[]{
    \includegraphics[width=0.46\linewidth]{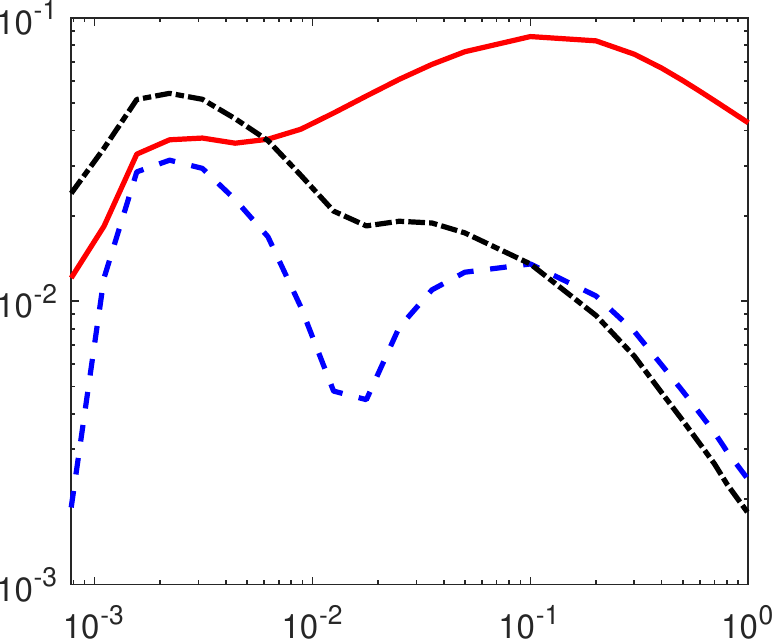}
  }  
  \hfill
  \caption[]
  {The relative error of various different approximations of numerical solution of equation \eqref{eq:AdvectionDiffusionEquationNon} with the flow $v (y,t)=\cos \pi y$, $\mathrm{Pe}=1$ and initial condition $T_I= \left( \sqrt{2 \pi } \sigma \right)^{-1}\exp \left(-\frac{x^{2}}{2\sigma^{2}}\right), \sigma=1/20$. The red solid, blue dashed and black dot dashed curve represent the relative error of approximation \eqref{eq:approximationTo1v0}, \eqref{eq:approximationTo1v1} and \eqref{eq:approximationTo2v1}. Panel (a) $\theta_1$ and $\theta_{2}$ are provided in \eqref{eq:ApproximationCos1O2NonI}. Panel (b) $\theta_1$ and $\theta_{2}$ are provided in   \eqref{eq:ApproximationCos1O2ZeroI} }
  \label{fig:ApproximationCos1O2}
\end{figure}

To verify the validity of the approximation \eqref{eq:approximationTo2v1},  we compare it with the numerical solution of equation \eqref{eq:AdvectionDiffusionEquationNon} with the flow $v (y,t)= \cos \pi y$. Solving equation \eqref{eq:centerCellProblemOrder1} and \eqref{eq:centerCellProblemOrder2}, we have
\begin{equation}\label{eq:ApproximationCos1O2NonI}
\begin{aligned}
&\theta_{1}= -\mathrm{Pe}\frac{\cos \pi y}{\pi^{2}}, \quad \theta_{2}=\frac{\mathrm{Pe}^2 \cos (2 \pi  y)}{8 \pi ^4}.
\end{aligned}
\end{equation}
To fit the initial condition $T_I$, we can impose the initial condition $\theta_{1} (y,0)=\theta_{2} (y,0)=0$ and obtain the time-dependent solutions,
\begin{equation}\label{eq:ApproximationCos1O2ZeroI}
\begin{aligned}
&\theta_{1}= -\mathrm{Pe}\frac{\cos \pi y}{\pi^{2}} \left( 1- e^{-\pi^{2}t} \right), \quad \theta_{2}=\frac{\mathrm{Pe}^2 \cos (2 \pi  y)}{8 \pi ^4} \left( 1- e^{-4\pi^{2}t} \right).
\end{aligned}
\end{equation}
Figure \ref{fig:ApproximationCos1O2} shows the relative error of various different approximations. The numerical solution is obtained via the method described in detail in appendix \ref{sec:appendixNumericalMethod}. We have three observations. First, the formula \eqref{eq:approximationTo1v1} and \eqref{eq:approximationTo2v1} retaining cross-sectional variation provide more accurate approximation than \eqref{eq:approximationTo1v0}. Second, we can see that the second order approximation \eqref{eq:approximationTo2v1} has smaller error than the first order approximation \eqref{eq:approximationTo1v1} at larger time.  We expect this difference will be more pronounced at longer times.  Third, if we impose the initial condition on $\theta_{1}$ and $\theta_{2}$, then we obtain a more accurate approximation at earlier stage.

Next, we consider the case $a_{3} \neq 0$. When the initial condition is $T_{I} (x,\mathbf{y})=\delta (x)$, the integral representation of the solution is 
\begin{equation}\label{eq:solutionBKDV}
\begin{aligned}
  \bar{T} (x,t) =&\frac{1}{2\pi} \int\limits_{-\infty}^{\infty} \exp(-b_{2} k^{2}-\mathrm{i}b_{3} k^{3} +\mathrm{i} x k)\mathrm{d} k \\
  =&\frac{1}{\pi} \int\limits_0^{\infty} \exp(-b_{2} k^{2})\cos( -b_{3} k^{3} +x k)\mathrm{d} k,  \\
\end{aligned}
\end{equation}
where $b_{3}=\int\limits_0^t a_{3} (s)\mathrm{d} s$. We are interested in the asymptotic expansion of solution \eqref{eq:solutionBKDV} at long times. It is a hard task for a very general time varying flow. Therefore, we restrict our attention to the case where $a_{2}, a_{3}$ are constant. For some time varying flows, we can approximate $a_{2}, a_{3}$ with their time average at long times, for example, periodic time-varying flow. Hence, the asymptotic expansion we derived in the section also applies to these cases. 

If $x \ll t$ and $t\rightarrow \infty$, the integrand in equation \eqref{eq:solutionBKDV} is localized around $k=0$. Hence, we have the approximation
\begin{equation}\label{eq:order2HermitePolymialRpresentation}
\begin{aligned}
  \bar{T} (x,t)= &\int\limits_{-\infty}^{\infty} \left( 1-\mathrm{i}a_{3}k^{3}t + \frac{(-\mathrm{i}a_{3} k^{3}t)^{2}}{2} \right) \exp(-a_{2} k^{2}t+\mathrm{i} x k)+\mathcal{O} (t^{-2})\\
  =& \left( 1- \frac{a_{3}  }{2^{3}a_{2}^{\frac{3}{2}} t^{\frac{1}{2}}}H_{3} \left( \frac{x}{2 \sqrt{a_{2}t }} \right)+  \frac{a_{3}^{2} }{2^{7} a_{2}^{3}t}  H_{6} \left( \frac{x}{2 \sqrt{a_{2}t}} \right) \right)\frac{\exp \left( \frac{-x^{2}}{4 a_{2}t} \right)}{\sqrt{4\pi a_{2}t}} +\mathcal{O} (t^{-2}),
\end{aligned}
\end{equation}
where $H_n$ is the degree $n$ Hermite polynomial associated with the weight function $e^{-x^{2}}$.
The approximation \eqref{eq:order2HermitePolymialRpresentation} is identical to the Hermite polynomial representation proposed in equation (5.7) in \cite{smith1982gaussian}.

\subsection{Improvements compared with previous studies}
We remark that there are two subtle differences compared with the previous studies \cite{mercer1990centre,mercer1994complete,marbach2019active}.  First, the previous studies  made not only the ansatz of the expansion of $T$, but also the expansion of $\bar{T}$. Therefore, the recursive equations involve not only  $\theta_n$, but also  the coefficients in the expansion of $\bar{T}$.  Here, we avoid making the expansion ansatz for $\bar{T}$ by utilizing equation \eqref{eq:ADEMeanEvolution}, which simplifies the calculation of $\theta_{n}$.

Second, in the previous studies, the center manifold are assumed to be time independent.  Hence, the equation for the auxiliary function $\theta_{1}$ derived in \cite{mercer1990centre,mercer1994complete,marbach2019active} takes the form
\begin{equation}\label{eq:centerCellProblemOrder1a}
\begin{aligned}
-\Delta_{\mathbf{y}} \theta_1=-\mathrm{Pe} (u-\bar{u}),\\
\end{aligned}
\end{equation}
in which the time derivative term doesn't appear. We think the justification is that the flow $u (\mathbf{y},t)$ varies slowly in time so that the time derivative term is negligible. However, we think this assumption limits many naturally arising applications. Let's consider a simple example, $\Omega=[0,1]$, $u=-  e^{\mathrm{i} \omega t} \cos \pi y $. The solution of equation \eqref{eq:centerCellProblemOrder1} is $\frac{ e^{\mathrm{i} \omega t} \cos \pi y}{\pi^{2}+\mathrm{i}\omega }$, while the solution of  equation \eqref{eq:centerCellProblemOrder1a} is $\frac{ e^{\mathrm{i} \omega t} \cos \pi y}{\pi^{2} }$. The only difference between them is the wavenumber $\omega$ in the denominator, which yields a $\mathcal{O} (\omega)$ difference. Hence, for small wavenumber $\omega$, the two solutions are close. However, for any fixed $\omega$, the corresponding approximations of the solute distribution $T$ diverge at long times, due to the variances having different growth rates. 
Recall that the variance,$\mathrm{Var} ( \bar{T})= 2 ( 1-\mathrm{Pe}\overline{v\theta_{1} } ) t+ \mathcal{O} (1)$,  grows linearly at long times. The difference between the two variances arising from the two different cell problems accumulates and becomes an $\mathcal{O} (1)$ difference at the frequency time scale $\mathcal{O} (\frac{1}{\omega})$. Since the solute distribution is characterized by the variance, the $\mathcal{O} (1)$ difference between variances implies an $\mathcal{O} (1)$ difference in the distributions at that time. Moreover, this difference in distributions will keep increasing as time increases. 
Hence, we conclude that equation \eqref{eq:centerCellProblemOrder1a} should only be used with a slow varying flow and before the frequency time scale.  In addition, this can be considered as an example of non-commutating limits. 

We know the center manifold becomes a good approximation if the exponential correction is small, i.e., after the diffusion time scale $L^{2}/\kappa$. If the frequency time scale is less than the diffusion time scale, then equation \eqref{eq:centerCellProblemOrder1a} is invalid for all time. That certainly limits the application of the result based on equation \eqref{eq:centerCellProblemOrder1a}. \cite{marbach2019active,masri2019reduced} adopted equation \eqref{eq:centerCellProblemOrder1a}  to study dispersion induced by pulsating flows. One of their applications is to blood flows.  Consider the following practical example. The typical frequency time scale in the human blood vessel is $1$s (60 heartbeats per min). The sodium chloride ($\kappa \approx 1.6*10^{-5} cm^{2}/s$ in water \cite{guggenheim1954diffusion}) diffuses cross the blood vessel with diameter $0.2$ mm takes around $25$ s. In this case, the result based on \eqref{eq:centerCellProblemOrder1a} is unlikely valid.

\begin{figure}
  \centering
\subfigure{
    \includegraphics[width=0.45\linewidth]{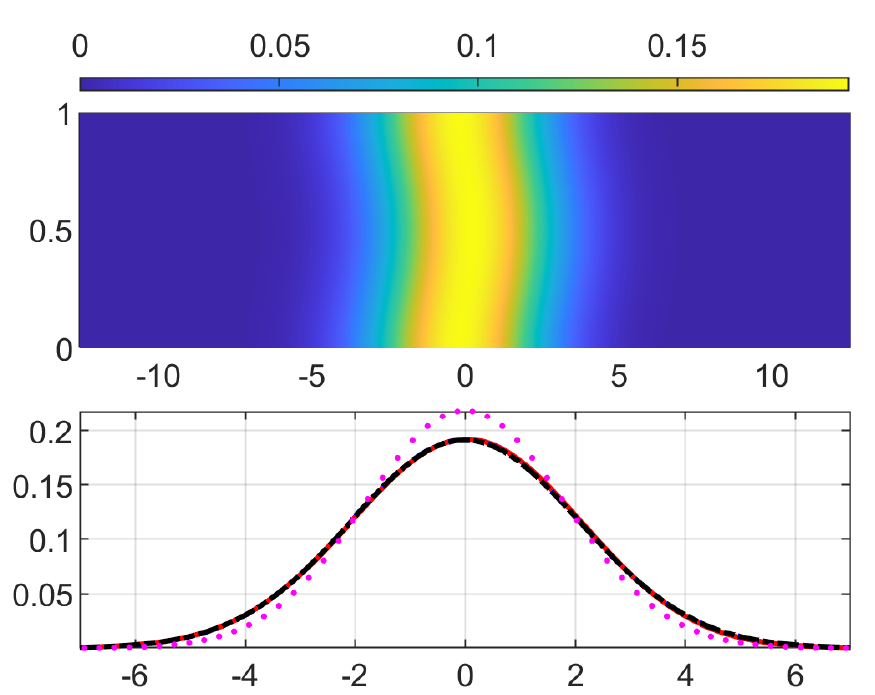}
}
\subfigure{
    \includegraphics[width=0.45\linewidth]{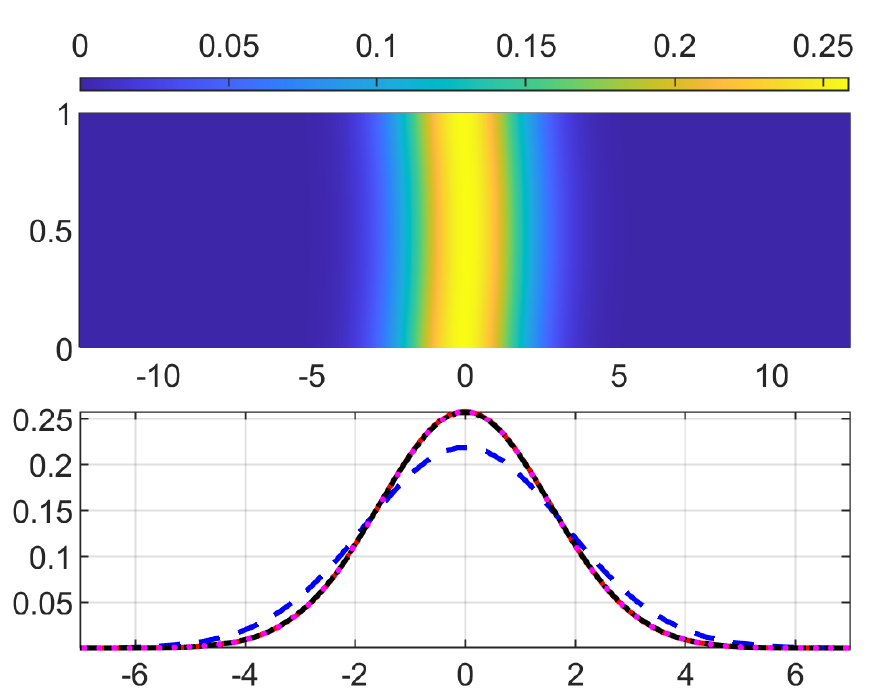}
}

  \hfill
  \caption[]
  {First row: Numerical solution of equation \eqref{eq:AdvectionDiffusionEquationNon} at $t=1$ with the shear flow $u (y,t)=\left( \cos \omega t \right) {y (1-y)}/{2}$, $\mathrm{Pe}=200$ and initial condition $T_I= \left( \sqrt{2 \pi } \sigma \right)^{-1}\exp \left(-\frac{x^{2}}{2\sigma^{2}}\right), \sigma=1/40$, where $\omega=\pi/5$ in left panel, and $\omega=20 \pi$ in right column. Second row: red curve is the cross sectional average of the numerical solution. The blue dash curve is the solution of equation \eqref{eq:mercer}. The black dot dash curve is the solution of equation \eqref{eq:centerEffectiveExample1}. The purple dot curve is the solution of equation \eqref{eq:centerEffectiveExample1average}.   }
  \label{fig:plotRoberts}
\end{figure}

To demonstrate the validity of our analysis, we solve equation \eqref{eq:AdvectionDiffusionEquationNon} numerically and present the results in figure \ref{fig:plotRoberts}. For the time varying shear flow $u (y)=\xi (t)y (1-y)/{2}$, \cite{mercer1990centre} derived the effective equation 
\begin{equation}\label{eq:mercer}
\begin{aligned}
\partial_{t}\bar{T}+ \frac{\mathrm{Pe} \xi (t)}{12}\partial_{x}\bar{T}= & \left( 1+\frac{ \mathrm{Pe}^{2}\xi (t)^{2}}{30240} \right)\partial_{x}^{2}\bar{T}. \\
\end{aligned}
\end{equation}
If $\xi (t)=\cos \omega t$, the solution of equation  \eqref{eq:centerCellProblemOrder1}  is
\begin{equation}
\begin{aligned}
 \theta=&\mathrm{Pe}\sum\limits_{n=1}^{\infty} \frac{(-1)^n+1}{ \pi ^2 n^2}\cos n\pi y \left( \frac{ \omega  \sin (t \omega )}{\pi ^4 n^4+\omega ^2}+\frac{\pi ^2 n^2  \cos (t \omega )}{\pi ^4 n^4+\omega ^2} -\frac{\pi ^2 n^2 \text{Pe} e^{-\pi ^2 n^2 t}}{\pi ^4 n^4+\omega ^2}\right)\\
\end{aligned}
\end{equation}
Hence the effective equation \eqref{eq:centerEffectiveOrder1} derived by time-dependent center manifold theory is
\begin{equation}\label{eq:centerEffectiveExample1}
\begin{aligned}
\partial_{t}\bar{T}+\frac{\mathrm{Pe} \cos \omega t}{12}\partial_{x}\bar{T}= & \left( 1+ \mathrm{Pe}^{2}\sum\limits_{n\in \text{even}^{+}}^{\infty}\frac{2\cos ^2(t \omega )}{\pi ^2 n^2 \left(\pi ^4 n^4+\omega ^2\right)}+ \frac{\omega  \sin (2 t \omega )}{\pi ^4 n^4 \left(\pi ^4 n^4+\omega ^2\right)}   \right)\partial_{x}^{2}\bar{T},
\end{aligned}
\end{equation}
where we neglect the exponential term in the solution of equation \eqref{eq:centerCellProblemOrder1}. When $t \gg \frac{1}{\omega}$, we could approximate the series in the effective equation by its time average
\begin{equation}\label{eq:centerEffectiveExample1average}
\begin{aligned}
\partial_{t}\bar{T}+\frac{\mathrm{Pe} \cos \omega t}{12}\partial_{x}\bar{T}= & \left( 1+ \mathrm{Pe}^{2}\left( \frac{1}{24 \omega ^2}-\frac{\sin \left(\frac{\sqrt{\omega }}{\sqrt{2}}\right)-\sinh \left(\frac{\sqrt{\omega }}{\sqrt{2}}\right)}{4 \sqrt{2} \omega ^{5/2} \left(\cos \left(\frac{\sqrt{\omega }}{\sqrt{2}}\right)-\cosh \left(\frac{\sqrt{\omega }}{\sqrt{2}}\right)\right)} \right)  \right)\partial_{x}^{2}\bar{T}.
\end{aligned}
\end{equation}
which is identical to the result of standard homogenization theory \cite{ding2021enhanced}.  Equation \eqref{eq:centerEffectiveExample1average} is simpler and performs as well as equation \eqref{eq:centerEffectiveExample1} at sufficiently large time scales. Of course, at intermediate times scales or in the case with irregular fluctuating flows, equation \eqref{eq:centerEffectiveExample1} performs better.

Figure \ref{fig:plotRoberts} shows the comparison of the numerical solution and  various different approximations at diffusion time scale $t=1$. The left column shows the result for a small frequency, $\omega=\pi/5$.  The cross-sectional average of the numerical solution, the solution of effective equations \eqref{eq:mercer} and \eqref{eq:centerEffectiveExample1} are almost indistinguishable. Recall that the standard homogenization result \eqref{eq:centerEffectiveExample1average} requires $t\gg \mathcal{O} (\frac{1}{\omega})$. As we expected, the standard homogenization result on this timescale is substantially worse than both center manifold results.  Alternatively, at higher frequency, with $\omega=20\pi$,  \eqref{eq:mercer} performs visibly worse than both standard homogenization \eqref{eq:centerEffectiveExample1average} as well as the time-dependent center manifold results \eqref{eq:centerEffectiveExample1}.  These observations from the numerical simulation are consistent with our previous theoretical analysis.  

\section{ Time varying random flows}\label{sec:randomflow}
Most studies of Taylor dispersion focused on periodic time varying flows, fewer studies have addressed irregularly fluctuating flows and even random flows.
In this section, we will show that the theory we developed in the previous section can be applied to time varying random flows. Moreover,  for random flows involving a white noise process or renewing processes, we show that the effective diffusivity is deterministic at long times.

This is also inspired by our work \cite{ding2020ergodicity} which studied the advection-diffusion equation with the shear flow $(v (y,\xi (t)),0)$ where $\xi (t)$ is a stationary Ornstein-Uhlenbeck (OU) process in  parallel-plate channels enforcing the no-flux boundary conditions. We derived the effective equation at long times via analyzing the $N$-point correlation function of the random scalar field. The analysis shows an interesting result that, in this random system, the effective diffusivity is deterministic at long times.

First, we consider the case that $\xi (t)$ is a Gaussian white noise process which is a zero-mean, Gaussian random process whose correlation function is given by $\left\langle \xi(t)\xi(s) \right\rangle=\delta (t-s)$. The center manifold approach is clearly valid for a smoothly varying velocity field. As for the Gaussian white noise which is non-differentiable, we can consider a sequence of function which converges to the white noise process.
The Wong-Zakai theorem states \cite{wong1965convergence,eugene1965relation,hairer2015wong} that the convergence of a process to white noise process yields the convergence of the systems driven by them. That justifies the application of the center manifold approach in the non-differentiable case involving white noise.

By utilizing the ergodicity of the white noise process and equation \eqref{eq:centeManifoldrEffectiveDiffusivity}, we obtain the effective diffusivity 
\begin{equation}\label{eq:effectiveWhiteTimeDiff}
\begin{aligned}
  \kappa_{\mathrm{eff}}&=1+ \mathrm{Pe}^{2}\left(\frac{1}{\left| \Omega \right|}\int\limits_{\Omega} u^{2} (\mathbf{y}) \mathrm{d} \mathbf{y} - \left(\frac{1}{|\Omega|} \int\limits_{\Omega} u (\mathbf{y}) \mathrm{d} \mathbf{y} \right)^{2}  \right).\\
\end{aligned}
\end{equation}
 Equation \eqref{eq:effectiveWhiteTimeDiff} is identical to equation (18) in \cite{ding2020ergodicity} which is derived via the analysis of $N$-point correlation function and Hausdorff moment problem. For the system with the random flows, in general, one has to repeat the experiment with different realizations of the flows to obtain the properties of the passive scalar via ensemble average. However, the deterministic diffusivity presented in equation \eqref{eq:effectiveWhiteTimeDiff} implies that one need only observe a {\em single} realization of the passive scalar to access some measurable quantities.

Second, we switch our attention to a class of stochastic flows with finite correlation time. 
Consider a shear flow  takes the form $(A(t)\xi(t)u(\mathbf{y}),0)$, where $\xi(t)$ is periodic function with a base frequency $\omega$, or equivalent, a period $L_{t}=\frac{2\pi}{\omega}$. $A(t)$  is a piecewise-constant zero-mean random function of time,
\begin{equation}
\begin{aligned}
A(t)= A_{n}, \quad nL_{t}\leq t < (n+1)L_{t},\;n \in \mathbb{Z},
\end{aligned}
\end{equation}
where $A_n$ is an independent and identically distributed random variable with zero mean and finite variance. This type of flow is in the class of renewing (renovating, innovation) flows, that is, flows that decorrelate completely in a finite time, taken here to be the period $L_{t}$. Therefore, it is a good approximation to a stationary process with a finite correlation time. It has wide applications in the study of the dynamo \cite{bhat2015fluctuation,zel1984kinematic} as well as in study of the intermittency of passive-scalar decay\cite{vanneste2006intermittency,haynes2005controls}. For this type of flow, the closed evolution equation for the statistical moment is unknown. Hence, the Hausdorff moment problem approach proposed in \cite{ding2020ergodicity} for rigorously studying the white noise flow case doesn't apply to this case. However, we could apply the center manifold approach to near rigorously derive the effective equation at long times.

In this case, the time averaged diffusion coefficients is
\begin{equation}\label{eq:effectiveDiffusivityRenewPeriodic}
\begin{aligned}
 \kappa_{\mathrm{eff}}&= 1 + \lim\limits_{t\rightarrow \infty} \frac{\mathrm{Pe}^{2}}{t} \sum\limits_{n=1}^{\infty} \left\langle u,\phi_{n} \right\rangle^{2} \int\limits_{0}^{t} e^{-\lambda_{n} s}\xi (s)A (s)\int\limits _0^se^{\lambda_{n} \tau} \xi(\tau)A (\tau) \mathrm{d} \tau \mathrm{d} s.\\
\end{aligned}
\end{equation}
We can further simplify this formula by utilizing the property of the renewing process. 

To take advantage of the periodicity, we tessellate the integral region by squares. The double integral in equation \eqref{eq:effectiveDiffusivityRenewPeriodic} becomes
\begin{equation}\label{eq:ArisTA2RenewPeriodic1}
  \begin{aligned}
   &  \sum\limits_{m_1=0}^{\lfloor \frac{t}{L_{t}}\rfloor-1}A_{m_{1}}^{2}\int\limits_{ m_{1} L_{t}}^{ (m_{1}+1) L_{t}}\int\limits _{m_{1}L_{t}}^s e^{-\lambda_{n} (s-\tau)}\xi (s) \xi(\tau)\mathrm{d} \tau \mathrm{d} s +A_{\lfloor \frac{t}{L_{t}}\rfloor}\int\limits_{L_{t}\lfloor \frac{t}{L_{t}}\rfloor}^{t} \int\limits _{0}^s e^{-\lambda_{n} (s-\tau)}\xi (s) A (\tau)\xi(\tau)\mathrm{d} \tau \mathrm{d} s\\
&+ \sum\limits_{m_1=1}^{\lfloor \frac{t}{L_{t}}\rfloor-1} \sum\limits_{m_2=0}^{m_{1}}
A_{m_{1}}A_{m_{2}} \int\limits_{m_{1}L_{t}}^{(m_{1}+1)L_{t}}\int\limits_{m_{2}L_{t}}^{(m_{2}+1) L_{t}}
 e^{-\lambda_{n} (s-\tau)}\xi (s) \xi(\tau)\mathrm{d} \tau \mathrm{d} s 
\end{aligned}
\end{equation}
In fact, only the first term in equation \eqref{eq:ArisTA2RenewPeriodic1} grows linearly on time. Thus it has the dominant contribution at long times.  To demonstrate this point, we will show that the second and third term  are bounded in time. The second term is an integral over a bounded interval $s\in [L_{t}\lfloor \frac{t}{L_{t}}\rfloor, t]$. It is enough to show the integrand is a bounded function of $s$ on this interval. We have 
\begin{equation}
\begin{aligned}
& \int\limits _{0}^s e^{-\lambda_{n} (s-\tau)}\xi (s) A (\tau)\xi(\tau)\mathrm{d} \tau \\
=&A_{\lfloor \frac{t\omega}{2\pi}\rfloor}\int\limits _{\frac{2\pi}{\omega}\lfloor \frac{t\omega}{2\pi}\rfloor}^s e^{-\lambda_{n} (s-\tau)}\xi (s) \xi(\tau)\mathrm{d} \tau + \sum\limits_{m_{2}=0}^{\lfloor \frac{t\omega}{2\pi}\rfloor-1}A_{m_{2}} \int\limits_{\frac{2\pi m_{2}}{\omega}}^{\frac{2\pi(m_{2}+1)}{\omega}}
e^{-\lambda_{n} (s-\tau)}\xi (s) \xi(\tau)\mathrm{d} \tau \\
=&A_{\lfloor \frac{t\omega}{2\pi}\rfloor}\int\limits _{\frac{2\pi}{\omega}\lfloor \frac{t\omega}{2\pi}\rfloor}^s e^{-\lambda_{n} (s-\tau)}\xi (s) \xi(\tau)\mathrm{d} \tau + \sum\limits_{m_{2}=0}^{\lfloor \frac{t\omega}{2\pi}\rfloor-1}A_{m_{2}} e^{-\lambda_{n} \left( s-\frac{2\pi m_{2}}{\omega} \right)} \int\limits_{0}^{\frac{2\pi}{\omega}}
 e^{\lambda_{n}\tau}\xi (s) \xi(\tau)\mathrm{d} \tau, \\
\end{aligned}
\end{equation}
where both terms in the last step are bounded functions of $s$. Next, we consider the third term in equation \eqref{eq:ArisTA2RenewPeriodic1}. With rearranging the order of the double summation, we have
\begin{equation}
\begin{aligned}
&\sum\limits_{m_1=1}^{\lfloor \frac{t\omega}{2\pi}\rfloor-1} \sum\limits_{m_2=0}^{m_{1}}
A_{m_{1}}A_{m_{2}}e^{-\lambda_n \frac{2\pi (m_{1}-m_{2})}{\omega}} \int\limits_{0}^{\frac{2\pi }{\omega}}\int\limits_{0}^{\frac{2\pi}{\omega}}
e^{-\lambda_{n} (s-\tau)}\xi (s) \xi(\tau)\mathrm{d} \tau \mathrm{d} s \\
=&\sum\limits_{q=1}^{\lfloor \frac{t\omega}{2\pi}\rfloor-1} \left( e^{-\lambda_n \frac{2\pi q}{\omega}} \int\limits_{0}^{\frac{2\pi}{\omega}}\int\limits_{0}^{\frac{2\pi }{\omega}}
 e^{-\lambda_{n} (s-\tau)}\xi (s) \xi(\tau)\mathrm{d} \tau \mathrm{d} s  \sum\limits_{m_{1}=q}^{\lfloor \frac{t\omega}{2\pi}\rfloor-1}
A_{m_{1}}A_{m_{1}-q} \right) \\
\end{aligned}
\end{equation}
For a fixed $q$, $\sum\limits_{m_{1}=q}^{\lfloor \frac{t\omega}{2\pi}\rfloor-1}
A_{m_{1}}A_{m_{1}-q}\rightarrow E (A_{q}A_{0})+\mathcal{O} (1) =\mathcal{O} (1)$ almost surely   because of the law of large numbers. In addition, the summand decays rapidly as $q$ increases. Therefore, this summation is also bounded in time  almost surely. 

Now, we have the leading order approximation of equation \eqref{eq:effectiveDiffusivityRenewPeriodic} at long times,
\begin{equation}\label{eq:effectiveDiffusivityRenew}
\begin{aligned}
 \kappa_{\mathrm{eff}}&= 1 + \lim\limits_{t\rightarrow \infty} \frac{\mathrm{Pe}^{2}}{t} \sum\limits_{n=1}^{\infty} \left\langle u,\phi_{n} \right\rangle^{2} \int\limits_{0}^{\frac{2\pi}{\omega}}\int\limits _{0}^s e^{-\lambda_{n} (s-\tau)}\xi (s) \xi(\tau)\mathrm{d} \tau \mathrm{d} s \sum\limits_{m_1=0}^{\lfloor \frac{t\omega}{2\pi}\rfloor-1}A_{m_{1}}^{2}+\mathcal{O} (t^{-1})\\
&= 1 +  \frac{\mathrm{Pe}^{2}  \omega}{2\pi}\mathrm{Var} (A_{0})\sum\limits_{n=1}^{\infty} \left\langle u,\phi_{n} \right\rangle^{2}\int\limits_{0}^{\frac{2\pi}{\omega}}\int\limits _{0}^s e^{-\lambda_{n} (s-\tau)}\xi (s) \xi(\tau)\mathrm{d} \tau \mathrm{d} s. 
\end{aligned}
\end{equation}
where the second step follows the law of large numbers.

It is natural to compare the renewing flow $(A (t)\xi (t)u (\mathbf{y}),\mathbf{0})$ with its deterministic counterpart $(\mathrm{var} (A_0)\xi (t)u (\mathbf{y}),\mathbf{0})$, and ask the question which one induces a larger effective diffusivity. One may expect the random motion creates a larger dispersion. However, it is not always true. A counter example is the $\xi (t)= \cos t$, $\mathrm{Pe}=1$ and $\mathrm{Var} (A)=1$, where the effective diffusivity induced by the renewing flow is $\kappa_{\mathrm{eff,r}} \approx 1.3993$, while the effective diffusivity induced by its deterministic counterpart is $\kappa_{\mathrm{eff,d}} \approx 1.4124$.

Interestingly, if we only consider the continuous renewing flow, then we have $\kappa_{\mathrm{eff,r}}\geq  \kappa_{\mathrm{eff,d}}$. The continuity of $A (t)\xi (t)$ implies $\xi (0)=\xi (L_{t})=0$. Hence, $\xi (t)$ admits a sine expansion $\xi (t)= \sum\limits_{k=1}^{\infty} c_k \sin k \omega t$.
Since the only difference in the effective diffusivity formula between the periodic in time case and the renewing process is the third integral in equation  \eqref{eq:ArisTA2RenewPeriodic1}, it is enough to establish that this third term is non-positive.
We have
\begin{equation}
  \begin{aligned}
 &\int\limits_{m_{1}L_{t}}^{(m_{1}+1)L_{t}}\int\limits_{m_{2}L_{t}}^{(m_{2}+1) L_{t}}
 e^{-\lambda_{n} (s-\tau)}\xi (s) \xi(\tau)\mathrm{d} \tau \mathrm{d} s \\
 =&e^{-\lambda_{n}L_{t} (m_{1}-m_{2})}\int\limits_{0}^{\frac{2\pi}{\omega}}\int\limits _{0}^{\frac{2\pi}{\omega}} e^{-\lambda_{n} (s-\tau)} \xi (s) \xi(\tau)\mathrm{d} \tau \mathrm{d} s \\
 =&e^{-\lambda_{n}L_{t} (m_{1}-m_{2})} \sum\limits_{k_{1},k_{2=1}}^{\infty}c_{k_{1}}c_{k_{2}} C_{k_{1},k_{2}}.
\end{aligned}
\end{equation}
where $ C_{k_{1},k_{2}}=\int\limits_{0}^{\frac{2\pi}{\omega}}\int\limits _{0}^{\frac{2\pi}{\omega}} e^{-\lambda_{n} (s-\tau)}  \sin (k_{2} \omega s)\sin (k_{1} \omega \tau)  \mathrm{d} \tau \mathrm{d} s$.  It is enough to show the (possibly infinite) matrix $C$ is semi-negative definite. In fact, we have
\begin{equation}
\begin{aligned}
C_{k_{1},k_{2}}= &-4\omega ^2 \sinh ^2\left(\frac{\pi ^3 n^2}{\omega }\right) \frac{ k_1 }{k_1^2 \omega ^2+ \lambda_{n}^{2}}\frac{ k_2 }{k_2^2 \omega ^2+\lambda_{n}^2}. \\
\end{aligned}
\end{equation}
Hence, for any $n$, $C$ is a rank one matrix with only one negative eigenvalue, which implies $C$ is semi-negative definite. Now, we finished the proof of $\kappa_{\mathrm{eff,r}}\geq  \kappa_{\mathrm{eff,d}}$ for the continuous renewing flow.

\begin{figure}
  \centering
    \includegraphics[width=1\linewidth]{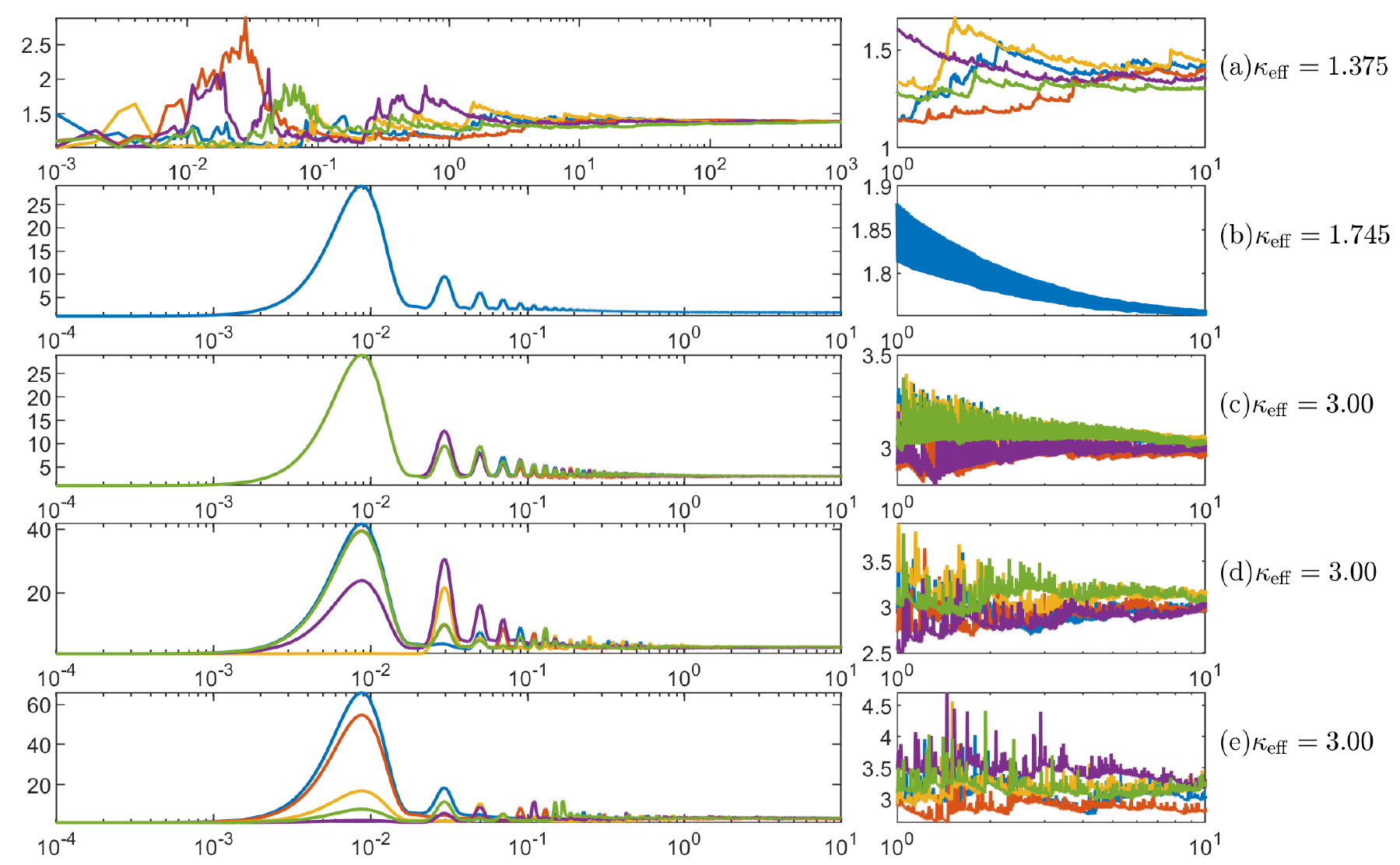}
  \caption{$\frac{\mathrm{Var} (\bar{T})}{2t}$ as a function of time for $5$ independent flow realizations and different random flows.  Note the convergence of this quantity to the deterministic effective diffusivity given in equation \eqref{eq:effectiveDiffusivityDefinition}. We compute $\kappa_{\mathrm{eff}}$ by equation \eqref{eq:effectiveWhiteTimeDiff} for panel (a), by equation \eqref{eq:centeManifoldrEffectiveDiffusivity} for panel (b), by equation \eqref{eq:effectiveDiffusivityRenew} for panel (c, d, e) and report three significant digits of the effective diffusivity to the right of each panel.  Pictures in the right column are simply zoom-in of  pictures in the left column  at a larger time scale. }
  \label{fig:differentRealization}
\end{figure}

To verify our theoretical results regarding the deterministic effective diffusivity, we solve equation \eqref{eq:AdvectionDiffusionEquationNon} with different shear flows by using the forward Monte-Carlo method described in \cite{ding2021enhanced}. The computational domain is $(x,y) \in \mathbf{R} \times [0,1]$. The time step size is $10^{-3}$. The total number of the random walkers is  $2\times10^6$. We divide a simulation into $400$ parallel jobs on UNC's Longleaf computing cluster. 
The shear flow takes the form $ v(y,t)=A (t)\sin t (y-1/2)$. In panel (a), $A (t)$ is a white noise process. In panel (b), $A (t)=1$. In panel (c, d, e), $A (t)$ is a renewing process with a coin-toss random variable taking values plus or minus one with equal probability, a uniform distributed random variable on $[-\sqrt{3}, \sqrt{3}]$  and a standard Gaussian distributed random variable respectively. We plot $\frac{\mathrm{Var} (\bar{T})}{2t}$ as a function of time for $5$ independent flow realizations and  different shear flows in figure \ref{fig:differentRealization}. The curves with the same color are generated with the same seed from the same random number generator.

From figure \ref{fig:differentRealization}, we have four1 observations. First, in panel (a,c,d,e), all curves fluctuate randomly at the earlier stage but converge at later times to a deterministic effective diffusivity $\kappa_{\mathrm{eff}}$ given by equation \eqref{eq:effectiveDiffusivityRenew}. Second, since all distributions in panel (c, d, e) have the same unit variance, all renewing flows induce the same effective diffusivity at long times.  Third, comparing panel (b) and panel (c, d, e), we can see that renewing random flows induce a larger effective diffusivity than their deterministic counterpart, as just proven above. Fourth, from the right column of figure  \ref{fig:differentRealization}, we can see that if the distribution of $A (t)$ has a heavier tail, then  $\frac{\mathrm{Var} (\bar{T})}{2t}$ takes a longer time to converge to the theoretical limit.

\subsection{Invariant measure}

\begin{figure}
  \centering
  \subfigure{
    \includegraphics[width=0.46\linewidth]{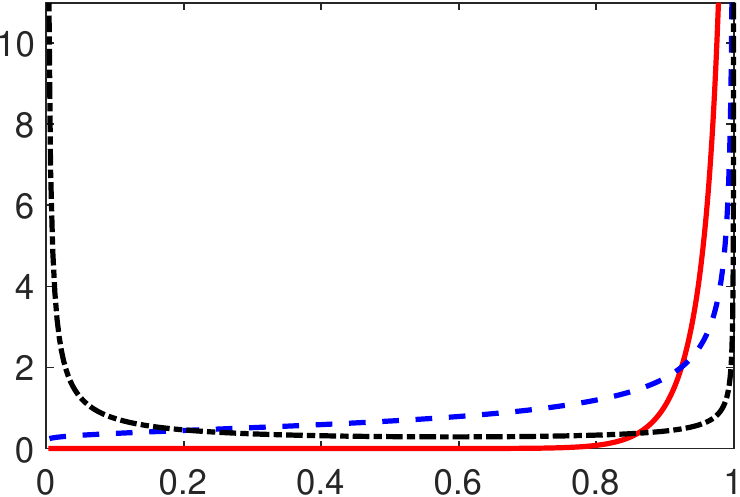}
  }
  \hfill
  \caption[]
  { The invariant measure $f_{\tilde{T}} (z)$  in equation \eqref{eq:invariantMeasureD1} for different parameters $\beta$. The red solid curve, blue dashed curve and black dot dash curve stands for the case $\beta=1/20, 1, 20$, respectively. $f_{\tilde{T}} (z)$ changes from negatively-skewed to positively-skewed as $\beta$ increases. }
  \label{fig:DeterministicPdf}
\end{figure}

Equation \eqref{eq:approximationTo1v0} is an approximation of the scalar field at long times, which is a powerful tool to compute the invariance measure of the random field. When $v(y,t)=\xi (t)u (\mathbf{y})$ and $\xi (t)$ is the Gaussian white noise, equation \eqref{eq:approximationTo1v0} becomes
\begin{equation}
\begin{aligned}
&  \bar{T} (x,t) =  \frac{1}{\sqrt{4\pi \kappa_{\mathrm{eff}} }} \exp \left( \frac{-\tilde{x}^{2}}{ 4 \kappa_{\mathrm{eff}}t} \right) + \mathcal{O} (t^{-\frac{3}{2}}),\quad  \tilde{x}=x- \mathrm{Pe}\bar{u} B(t),\\
  &\kappa_{\mathrm{eff}}=1+ \mathrm{Pe}^{2}\left(\frac{1}{\left| \Omega \right|}\int\limits_{\Omega} u^{2} (\mathbf{y}) \mathrm{d} \mathbf{y} - \left(\frac{1}{|\Omega|} \int\limits_{\Omega} u (\mathbf{y}) \mathrm{d} \mathbf{y} \right)^{2}  \right). \\
\end{aligned}
\end{equation}
where $B (t)$ is the standard Brownian motion. Then we apply the inverse transform method (we refer reader to \cite{bronski2007explicit} for details) to obtain the invariant measure  of $\bar{T}$, i.e, the probability density function at long times, from the probability density function of $B (s)$. We consider the rescaling of $T$, $\tilde{T} (x,y,t)=\sqrt{4\pi\kappa_{\mathrm{eff}} t} T$. Without loss of generality, we focus on the scalar at point $x=0, y=0$, i.e., $\tilde{T}(0,0,t)$. Thus, the invariant measure is
\begin{equation}\label{eq:invariantMeasureD1}
\begin{aligned}
f_{\tilde{T}} (z)=\frac{z^{\frac{1}{\beta }-1}}{\sqrt{-\pi \beta  \log (z)}},\quad z\in[0,1],
\end{aligned}
\end{equation}
where  $\beta= \frac{\mathrm{Pe}^2 \bar{u}^2 v(t)}{2 t \kappa _{\text{eff}}} =\frac{\mathrm{Pe}^2 \bar{u}^2 }{2  \kappa _{\text{eff}}}+\mathcal{O} (t^{-1})$ and $v(t)$ is the variance of $\int\limits_0^t \xi (s)\mathrm{d}s$. $f_{\tilde{T}} (z)$ always has the logarithmic singularity at $z=1$. It is continuous at $z=0$ when $\beta\leq 1$, and singular when $\beta>1$ (see figure \ref{fig:DeterministicPdf}). This property of the distribution implies that when the strength of the input random signal exceeds some certain threshold, the value of scalar is more likely to be zero.  As a result of that, the distribution changes from negatively-skewed to positively-skewed as $\beta$ increases. 

\section{Conclusion and Discussion}

We have studied the long time behavior of an advection-diffusion equation
with a general time-varying (including random) shear flow imposing no-flux boundary
conditions on channel walls using center manifold theory.  Our study extends and improves prior work of \cite{mercer1990centre,mercer1994complete,marbach2019active} to properly incorporate general time variation into the effective theory.  Comparisons with full simulations document conditions when this improved approach gives a better approximation, and also illustrates situations in which standard homogenization does not perform on finite timescales.  Convergence studies illustrate how the accuracy of the different approximations.  Armed with this improved time varying center manifold theory, we derived new effective equations for random shear flows involving both white in time statistics, as well as more correlated renewing flows.  For white in time, these predictions agree with our prior work \cite{ding2020ergodicity}, which forecast a deterministic effective diffusivity on long times.  For the case of renewing flows, less is known, and our current work also a deterministic effective diffusivity, with new explicit formulae.  These theories are demonstrated to be quantitatively accurate through Monte-Carlo simulations.  New conditions are derived which guarantee when the random renewing flow generates a larger effective diffusivity than its deterministic analog.  Lastly, using inverse transform method and the effective equations, we derived the invariant measure and study its  P\'{e}clet number dependence.

In this study, we only considered constant diffusivity. Future immediate areas of exploration include case with spatial variable dependent diffusivity or even concentration dependent diffusivity. A practical example concerns the shear-enhanced diffusion in colloidal suspensions explored in \cite{griffiths2012axial}. The nonlinearity in those system imposes challenges to the traditional method. We expect center manifold theory could overcome the difficulties.   Further, center manifold theory will apply nicely to study the mixing ability of time-varying flow in a non-flat channel to generalize the conclusion in \cite{rosencrans1997taylor}.

\section{Acknowledgements}
We acknowledge funding received from the following NSF Grant Nos.:DMS-1910824; and ONR Grant No: ONR N00014-18-1-2490. Partial support for Lingyun Ding is gratefully acknowledged from the National Science Foundation, award NSF-DMS-1929298 from the Statistical and Applied Mathematical Sciences Institute.

\section{Appendix}\label{sec:appendix}

\subsection{Numerical Method}\label{sec:appendixNumericalMethod}
In this section, we document details of the algorithm for the numerical simulation of equation \eqref{eq:AdvectionDiffusionEquation}. The computational domain is  $x\times y \in [-H,H]\times[0,L]$. When $H$ is large enough, we can assume there is a periodic boundary condition in the $x$-direction. Since there are non-penetration conditions in the $y$-direction, we perform the even extension in the $y$-direction to obtain the periodic condition on the extended domain.  Thus, we can use the standard Fourier spectral method to solve the advection-diffusion equation with periodic boundary conditions on the rectangular domain $[-H, H]\times[0, 2L]$.  In the dealiasing process at each time step, we apply the all-or-nothing filter with the two-thirds rule to the spectrum, that is, we set the upper one-third of the resolved spectrum to zero (see chapter 11 of the book \cite{boyd2001chebyshev} for details). 

The diffusion operator is stiff, which requires a very small time step size for the explicit method to ensure numerical stability.  In order to use a larger time step size and improve the efficiency, we adopt the implicit-explicit third-order Runge-Kutta method presented in table 6 in \cite{pareschi2005implicit}. In our application, we use the explicit Runge-Kutta method to integrate the advection part and use the implicit diagonal Runge-Kutta method to integrate the diffusion term. When the diffusivity is a constant, the diffusion operator is a diagonal matrix in the Fourier space. Thus, the implicit equation can be solved explicitly and efficiently. The implicit-explicit method is as efficient as the explicit method at each iteration while allows a much larger time step size. When the diffusivity is a function of spatial variables, the implicit part requires solving a linear system, which is expansive. Therefore, in this case or when the advection is dominant, we adopt the explicit 4th-order Runge-Kutta method as the time-marching scheme.

We also present the Butcher tableau of the explicit-implicit Runge-Kutta method in table \ref{tab:EXIMRK} here for convenience.  Unfortunately, \cite{pareschi2005implicit} only reported 13-14 significant digits of parameters $(\alpha,\beta,\eta)$ which are the key parameters defining the algorithm. That may potentially deteriorate the accuracy of double-precision floating-point based or even higher precision floating-point based algorithms. Hence, we documented the exact value for those parameters, $(\frac{9-\sqrt{57}}{6},\frac{9-\sqrt{57}}{24},\frac{-6+\sqrt{57}}{12})$. We also find another two groups of parameters that achieve the same convergence order and ensure the L-stable, $(1/2,1/8,0)$ and $(\frac{9+\sqrt{57}}{6},\frac{9+\sqrt{57}}{24}, \frac{-6-\sqrt{57}}{12})$.
\begin{table}[h]
   \centering
 \begin{tabular}{l|llll}
   0&0&0&0&0 \\
   0&0&0&0&0\\
   1&0&1&0&0   \\
   1/2&0&1/4&1/4&0\\
   \hline
    &0&1/6&1/6&2/3\\
 \end{tabular}
\quad
 \begin{tabular}{l|llll}
 $\alpha$& $\alpha$ &0&0&0\\
       0&-$\alpha$&$\alpha$&0&0\\
        1&0&1-$\alpha$&$\alpha$&0\\
  1/2&$\beta$&$\eta$&$1/2-\beta-\eta-\alpha$&$\alpha$\\
   \hline
   &0&1/6&1/6&2/3\\
 \end{tabular}
\caption{Butcher tableau for the Explicit (left) Implicit (right) L-Stable scheme, $(\alpha,\beta,\eta)$ could be $(1/2,1/8,0)$, $(\frac{9-\sqrt{57}}{6},\frac{9-\sqrt{57}}{24},\frac{-6+\sqrt{57}}{12})$ or $(\frac{9+\sqrt{57}}{6},\frac{9+\sqrt{57}}{24}, \frac{-6-\sqrt{57}}{12})$. } \label{tab:EXIMRK}
\end{table} 
% \begin{table}[h]
%  \centering
%  \begin{tabular}{l|llll||l|llll}
% \hline\hline
%    0&0&0&0&0&   $\alpha$& $\alpha$ &0&0&0\\
%    0&0&0&0&0&         0&-$\alpha$&$\alpha$&0&0\\
%    1&0&1&0&0&         1&0&1-$\alpha$&$\alpha$&0\\
%    1/2&0&1/4&1/4&0&  1/2&$\beta$&$\eta$&$1/2-\beta-\eta-\alpha$&$\alpha$\\
%    \hline
%     &0&1/6&1/6&2/3&    &0&1/6&1/6&2/3\\
%                                    \hline\hline
%  \end{tabular}
% \caption{Butcher tableau for the Explicit (left) Implicit (right) IMEXSSP3(4,3,3) L-Stable scheme} \label{tab:abbreviations}
% \end{table} 

 % Typically, we solve equation with the parameters $H=16$, $L=0.2$, and time increment $\Delta t= 0.005$ over $2000$ timesteps. The grid resolution is $2048 \times 257$ before the even extension and  $2048 \times 512$ after the extension.

\subsection{Lists of abbreviations}
See table \ref{tab:abbreviations}.

\begin{table}[h]
 \centering
 \begin{tabular}{l|l}
\hline
  Full Form & Abbreviation\\
\hline\hline    
Ornstein-Uhlenbeck &OU\\
Partial differential equation & PDE \\
Probability distribution function & PDF\\
Stochastic differential equation &SDE\\   
\hline
 \end{tabular}
\caption{Lists of abbreviations.} \label{tab:abbreviations}
\end{table}

\bibliographystyle{elsarticle-harv}
%\bibliography{bib/PassiveScalar,bib/matrix,bib/stochastic,bib/TaylorDispersion,bib/Asymptotic,bib/MyPaper,bib/StokesFlow,bib/Experiment,bib/SpectralMethod}  

%%%%%%%%%%%%%%%%%%%%%%%%%%%%%%%%%%%%%%%
%%%%%%%%%%%%%%%%%%%%%%%%%%%%%%%%%%%%%%%
%%%%%%%%%%%%%%%%%%%%%%%%%%%%%%%%%%%%%%%
%%%%%%%%%%%%%%%%%%%%%%%%%%%%%%%%%%%%%%%
%%%%%%%%%%%%%%%%%%%%%%%%%%%%%%%%%%%%%%%

\end{document}